\documentclass{mn2e}
\usepackage{graphicx}
\usepackage{refcount}
\usepackage{amssymb}
\title[Radio-quiet lensed quasars]
{Observations of radio-quiet quasars at 10mas resolution by use of gravitational lensing}
\author[Jackson et al.]
{Neal Jackson$^{1}$, Amitpal S. Tagore$^{1}$, Carl Roberts$^{1}$, Dominique Sluse$^{2,3}$, \\
\mbox{}\\
{\rm \LARGE Hannah Stacey$^{1}$, Hector Vives-Arias$^{1,4}$,  Olaf Wucknitz$^{5}$, Filomena Volino$^{2}$}\\
\mbox{}\\
$^{1}$Jodrell Bank Centre for Astrophysics, School of Physics \& Astronomy, 
University of Manchester, Turing Building, Oxford Road, \\
Manchester M13 9PL, UK\\
$^{2}$Argelander Institut f\"ur Astronomie, Auf dem H\"ugel 69, 53121 Bonn, Germany\\
$^{3}$Institut d'Astrophysique et de G\'eophysique, Universit\'e de Li\`ege, All\'ee du 6 Ao\^ut 17, B5c, 4000 Li\'ege, Belgium\\
$^{4}$Departamento de Astronom\'{\i}a y Astrof\'{\i}sica, Universidad de Valencia, E-46100 Burjassot, Valencia, Spain\\
$^{5}$Max-Planck-Institut f\"ur Radioastronomie, Auf dem H\"ugel 69, 53121 Bonn, Germany\\
}
\begin{document}
\maketitle
\begin{abstract}
We present VLA detections of radio emission in four four-image gravitational lens systems with quasar sources: 
HS~0810+2554, RX~J0911+0511, HE~0435$-$1223 and SDSS~J0924+0219, and e-MERLIN observations of two of the systems. The first three are detected at a high level of 
significance, and SDSS~J0924+0219 is detected. HS~0810+2554 is resolved, allowing us for the first time to achieve 10-mas
resolution of the source frame in the structure of a radio quiet quasar. The others are unresolved or marginally resolved.
All four objects are among the faintest radio sources yet detected, with intrinsic flux densities in the range 1--5$\mu$Jy; such radio
objects, if unlensed, will only be observable routinely with the Square Kilometre Array. The 
observations of HS~0810+2554, which is also detected with e-MERLIN, strongly suggest the presence of a mini-AGN, with a radio core and milliarcsecond scale
jet. The flux densities of the lensed images in all but HE~0435$-$1223 are consistent with smooth galaxy lens models without
the requirement for smaller-scale substructure in the model, although some interesting
anomalies are seen between optical and radio flux densities. These are probably due to microlensing effects in the optical.
\end{abstract}


\large

\section{Introduction}

Strong gravitational lens systems, in which background sources are multiply imaged by foreground galaxies, are important 
for two main reasons. First, the lensing effect magnifies the background source; although the surface brightness is 
conserved, the area increases and we can observe background sources with typically 5--10 times better signal-to-noise than
without the lensing. Second, the lensing effect allows us to probe the mass distribution of the lensing galaxy,
on scales from the overall mass profile down to the scales of individual stars. General reviews of strong lensing and its 
applications relevant to this work are given by Kochanek (2004), Courbin, Saha \& Schechter (2002), Zackrisson \& Riehm 
(2010), Bartelmann (2010) and Jackson (2013). 

Lens systems in which the background source is a quasar were the first class of systems to be discovered, mostly in radio 
surveys (Walsh, Carswell \& Weymann 1979; Hewitt et al. 1988; Browne et al. 2003) but later in optical surveys (e.g. 
Wisotzki et al. 1993; Inada et al. 2003). Lens systems with optical, or ``radio-quiet'', quasars as the source now dominate 
the sample of strongly lensed quasars. The radio-selected sample is mostly composed of the 22 lenses from the Cosmic Lens All-Sky survey (CLASS; Myers 
et al. 2003; Browne et al. 2003) together with smaller samples from the MIT-Greenbank (MG) and southern surveys (e.g. Hewitt et al. 1992; 
Winn et al. 2002). This sample has not expanded significantly in the last 15 years, because of the difficulty -- in the era 
before the Square Kilometre Array -- of conducting more sensitive wide-field radio surveys at the necessary sub-arcsecond 
resolution for lens discovery.

In order to increase the size of the sample of quasar lenses observed at radio wavelengths, 
we can use the fact that all quasars are likely to have radio emission at some level. For example, White et al. (2007) 
performed a stacking analysis at the positions of radio-``quiet'' quasars from the FIRST 20-cm radio survey (Becker et al. 
1995) and found that quasars not detected at the 1-mJy level are likely, on average, to have radio flux densities which
fall with decreasing optical brightness to $\sim 100\mu$Jy at $I=20-21$.  
This is within reach of the new generation of telescopes such as the Karl G. Jansky Very Large Array (VLA) and the extended Multi-Element Remote Linked Interferometer (e-MERLIN). We began a programme (Wucknitz
\& Volino 2008, Jackson 2011) to investigate four-image quasar lens systems without current radio detections, and achieved 
a successful  detection, at around the 20-30$\mu$Jy level, of lensed images of the background quasar in the cluster lens 
system SDSS~J1004+4112, as well as detections of lensed images at a brighter level in the lens system RXJ~1131$-$1231.

There are three main motivations for radio observations of the radio emission from radio-``quiet''\footnote{In the rest of this
article, we drop the inverted commas, but it should be understood that radio-``quiet'' quasars are not radio silent.} quasar
lens systems. The first is that the emission mechanisms of radio-quiet quasars are not well understood, and in particular it 
is not clear how far the mechanisms which power radio-loud quasars -- accretion and the formation of a powerful jet close 
to a black hole -- also apply to radio-quiet quasars. In sources of intermediate flux density, jets characteristic of ordinary 
active nuclei are seen (Blundell \& Beasley 1998, Leipski et al. 2006) and the radio sources appear to be variable (Barvainis 
et al. 2005). Both of these observations suggest that the AGN paradigm applies to these objects. In fainter cases ($<$1~mJy 
at 1.4~GHz at redshifts of a few tenths) however, Condon et al. (2013) argue that star formation is the primary mechanism for 
the emission. This is inferred from an analysis showing that the radio luminosity function of QSOs  turns 
up sharply below $L_{\rm 1.4GHz}<10^{24}$~W$\,$Hz$^{-1}$ and suggesting that a second population emerges at these luminosities. It is 
also possible that a different emission mechanism is at work in the core of the sources. Blundell \& Kuncic (2007) suggest
the presence of optically-thin bremsstrahlung emission (although see Steenbrugge et al. 2011 for evidence against this view) 
and Laor \& Behar (2008) propose the possibility of emission from magnetically-heated coronae, rather than a classical AGN 
source. It is important to achieve high-resolution radio imaging of these sources in order to separate 
the possibilities. This is typically very difficult to achieve with current long-baseline interferometers in very faint objects.
However, the use of lensing magnification provides a way to detect otherwise unobservable objects.

The second motivation is that radio and optical observations are subject to different propagation effects. The main such 
effect in the optical is microlensing due to stars in the lens galaxy, which produces measurable flux changes 
because the size of the optical source
is very small. Repeated optical monitoring can reveal the flux density variations associated with the movement of the source
with respect to the microlensing caustic patterns (Irwin et al. 1989, Wisotzki et al. 1993; Burud et al. 2002; Poindexter,
Morgan \& Kochanek 2008; Blackburne et al. 2011; Mu\~noz et al. 2011). At radio wavelengths, the source is larger, and 
therefore much less susceptible to microlensing; hence, 
comparison between the two wavebands can reveal the effects of microlensing directly. In the radio, the only significant 
propagation effect is scattering by ionized media (Koopmans et al. 2003)\footnote{In principle, the size of a compact, 
synchrotron self-absorbed radio source decreases as the square root of the brightness, but this is unlikely to result in 
microlensing until the source is fainter than 1$\mu$Jy. Claims exist for radio microlensing (Koopmans et al. 2000) which
could also be explained by other propagation effects. In principle, free-free absorption is also possible (Mittal et al. 
2007) although the electron columns are likely to be too small in all but exceptional cases.}. This seems to be 
particularly noticeable in a few 
cases, such as CLASS~B0128+437 (Phillips et al. 2000, Biggs et al. 2004) but should in principle decrease strongly at higher 
radio frequencies. A corresponding problem at optical wavelengths is extinction by dust in the lensing galaxy, which can be 
used to learn about the properties of the dust if the intrinsic fluxes are known (e.g. Jackson, Xanthopoulos \& Browne 2000,
Eliasd\'ottir et al. 2006, Ostman et al. 2008). Radio wavelengths therefore provide an important input to any programme which
aims to disentangle the effects of substructure in the lens galaxy from those of microlensing and extinction.

The third motivation for radio observations of four-image gravitational lens systems is their potential to probe substructure
in the lens galaxies.  In principle, quasar lens systems are useful for probing small-scale structure within the lens
potential, down to 10$^6M_{\odot}$ and below (Mao \& Schneider 1998), because the flux density of the lensed image can be 
perturbed by small-scale mass structures close to the corresponding ray path. Such sub-galactic scale structures are important
predictions of Cold Dark Matter (CDM) models. Initially they were thought not to be present in required quantities in the Milky Way (Moore et al. 1999, 
Klypin et al. 1999). The situation is now less clear, as a population of faint Milky Way satellites have since been discovered 
(Belokurov et al. 2006; Zucker et al. 2006; Koposov et al. 2015). The Milky Way halo mass is a 
critical variable (Wang et al. 2012; Kafle et al. 2014) as the predicted halo population is sensitive to it. In lens systems, 
the flux density of the lensed images is particularly sensitive to small structures, because it depends on the second derivative 
of the lensing potential, as opposed to the image positions which depend on the first derivative. The usual evidence for a 
detection of substructure is therefore a set of image flux ratios which cannot be fit by smooth models. More particularly, 
four-image lenses in cusp configurations (where the source is close to the cusp of the astroid caustic) and fold configurations 
(resulting from the source being close to the caustic fold) give clear theoretical expectations for image flux ratios which 
must be obeyed by any smooth model. In cusp lenses, there are three close images and the middle image is expected to have 
the brightness of the sum of the outer two (Schneider \& Weiss 1992); in fold lenses, the two close images are expected to
have the same flux (Keeton, Gaudi \& Petters 2003; Congdon, Keeton \& Nordgren 2008). Because of optical microlensing, 
radio lens systems have traditionally been used for this work (Mao \& Schneider 1998, Fassnacht et al. 1999, Metcalf \& 
Zhao 2002, Metcalf 2005, Kratzer et al. 2011). 
The statistics of such objects were analysed by Dalal \& Kochanek (2002) who found a fraction of between 0.6\% and 7\% of mass 
in $10^6-10^9M_{\odot}$ substructures at the radius probed by the lensing. More recent theoretical work has used more realistic
treatment of lens galaxies via the use of numerical simulations (Bradac et al. 2004, Amara 2006, Macci\`o et al. 2006, Xu et 
al. 2009). The conclusions are generally that there is, if anything, an excess of substructure over what is predicted to exist 
in CDM (though see Metcalf \& Amara 2012, Xu et al. 2015). At the same time, analyses of individual lens systems have 
yielded constraints on substructures at the $\sim10^6M_\odot$ level (e.g. Fadely \& Keeton 2012).
The major problem in using quasar lenses to constrain substructure has been the lack of large enough samples of radio-loud
four-image lenses; the seven studied by Dalal \& Kochanek in 2002 have formed the sample for most subsequent investigations. 
Possible solutions to this problem include the use of mid-infrared fluxes (Chiba et al. 2005; Fadely \& Keeton 2011) assuming that
the mid-IR source is large enough not to be subject to microlensing (but see Sluse et al. 2013). An alternative approach is to 
use the narrow line region of quasars (Moustakas \& Metcalf 2003, Sugai et al. 2007, Nierenberg et al. 2014) which should also 
be large enough to be unaffected by microlensing, 
or submillimetre observations in the case of new lenses from Herschel and the SPT (Hezaveh et al. 2013). A further 
alternative is to perform lens reconstruction of systems with extended sources (Warren \& Dye 2003, Dye \& Warren 2005, 
Koopmans 2005, Vegetti et al. 2009, Vegetti et al. 2012). Cases of substructure detections have already been reported 
from these works, although the sensitivity is mainly towards the higher-mass end of the substructure mass function;
quasar lenses are thus likely to be usefully complementary to this method.

This work presents a continuation of a programme to detect and image faint radio sources in gravitational lens systems. Its aim 
is to increase the number of four-image lenses with detected radio fluxes, both to increase the sample sizes of quasar lens 
systems suitable for the investigation of sub-galactic scale substructures in the lens, and to begin the study of the 
very faint radio sources which are imaged by the foreground lens galaxies. Where necessary we use a standard flat cosmology with 
$\Omega_m=0.27$ and $H_0=68$kms$^{-1}$Mpc$^{-1}$.


\section{Sample and observations}

\subsection{Sample selection}

Our target sample includes all known gravitational lens systems with four lensed images, no detected radio emission above 
the $\sim$1~mJy level reached in large-scale sky surveys such as the FIRST and NVSS 1.4-GHz surveys (Becker et al. 1995, 
Condon et al. 1998), and with declination $>-20^{\circ}$ for accessibility to the VLA and e-MERLIN radio arrays.
There are 13 of these in current compilations such as the CASTLES (Kochanek et al. 1998) and Masterlens (Moustakas et al. 2012) 
lists, which represents a potential factor of 3 improvement in statistics if radio flux densities can be measured for
all of them. One of these objects, SDSS~J1004+4112, was already detected by Jackson (2011) using the VLA in the lower
resolution C-configuration. This is a wide-separation object (Inada et al. 2003a) resulting from the lensing action of a 
cluster, whose mass distribution is correspondingly more difficult to model. Most such objects, however, are lensed by 
individual galaxies; we have in this preliminary phase used the VLA (resolution $\sim$0\farcs3 at 5~GHz) to investigate the 
wider-separation objects within this sample. A further object, RXJ~1131-1231, was previously detected by Wucknitz \& 
Volino (2007) in archival VLA data, and subsequently with the VLA and MERLIN (although not with VLBI). 
Table \ref{observations} shows a summary of the lens systems observed (including, for completeness, SDSS~J1004+4112) 
together with the source and lens redshifts and other observational information.

\begin{table*}
\begin{tabular}{ccccccccc}
Object & $z_{\rm lens}$ & $z_{\rm source}$ & Separation & References &
    VLA Obs. date& Frequency &Exposure  & noise \\
      & & & /arcsec & & (2012) &/GHz & time/h & /$\mu$Jy/b \\
HE~0435$-$1223 & 0.46 & 1.69 & 2.5 & W02,M05,O06 & Oct 26, Nov 9& 5 & 6 & 3\\
HS~0810+2554 & ? & 1.50 & 0.9 & R02 & Oct 22, Nov 8,24& 8.4 & 7.5  &3\\
RX~J0911+0551 & 0.77 & 2.80 & 3.2 & B97,B98,K00 & Oct 31. Nov 6,24& 5 & 7.5 &3\\
SDSS~J0924+0219 & 0.39 & 1.52 & 1.8 & I03B,E06,O06 &Nov 5& 5 & 3 &3\\
SDSS~J1004+4112 & 0.68 & 1.73 & 14.6 & I03A & See Jackson 2011& 5 & 6 &3\\
\end{tabular}
\caption{Basic information for the systems studied, including
the redshifts of lens and source (where known), maximum separation of
the lensed images, observation time and frequency, and achieved noise
level in the maps. References are given to the discovery papers for
each lens, and to the sources for the measurements of the redshifts.
In each case the on-source integration time is approximately 75\% of
the total exposure time. Key to references: W02 = Wisotzki et al. 2002,
M05 = Morgan et al. 2005, R02 = Reimers et al. 2002, B97 = Bade et al.
1997, B98 = Burud et al. 1998, 
K00 = Kneib, Cohen \& Hjorth 2000, I03A = Inada et al. 2003a, 
I03B = Inada et al. 2003b, E06 = Eigenbrod et al. 2006, O06 = Ofek et 
al. 2006.}
\label{observations}
\end{table*}

\begin{table*}
\begin{tabular}{ccccc}
Object & Phase calibrator & Date & Exposure & Frequency \\
HS~0810+2554 & JVAS~0813+2435 & 31/03/2014  & 8h & 1287-1799MHz \\
RXJ0911+0551 & SDSS J0901+0448 & 26/04/2014  & 8h & 1287-1799MHz \\
\end{tabular}
\label{emerlin}
\caption{Details of the e-MERLIN observations of two of the sources.}
\end{table*}
%
%

\subsection{Observations and data reduction}

\subsubsection{VLA observations}

Objects were observed with a total bandwidth of 2~GHz in 16 IFs of 128 MHz over the frequency range 4488-6512 MHz (C-band). 
The exception was HS~0810+2554 which was observed at X-band, with a similar spectral arrangement over the frequency 
range 7988-10036 MHz, in order to achieve the resolution needed for the smaller spatial scale of this source. Integration 
times were generally a few hours (Table \ref{observations}) and observations were taken at various times during the autumn of
2012. Although the individual 3 or 1.5-hour observations were taken at different times, we do not detect variability in any 
case between individual epochs. All new observations were taken in A-configuration, which has a maximum baseline of 36~km and 
a theoretical resolution of 0\farcs35 at 5~GHz and 0\farcs22 at 8.4~GHz. Data were taken in  spectral-line mode, with 
2-MHz channels, although this was reduced in subsequent processing as only a small area of sky was required.

Nearby phase calibrators were observed at regular intervals, with a pattern of 1 minute on the calibrator and 5 minutes 
on source, in order to calibrate the instrumental and atmospheric phases. Sources of known flux density, either 3C138 or 
3C286, were observed in order to fix the flux density to the Baars et al. (1977) scale.

Data analysis was performed using the NRAO {\sc aips} package. The data were fringe fitted to remove instrumental delays 
using the phase calibrator observations, and a phase and amplitude solution was performed using the phase calibrator and a 
point source model. Maps were also made of the phase calibrator and used where necessary to iterate the phase and amplitude
calibration, and the flux density solution was transferred from the flux calibrators. The resulting calibration was then 
applied to the target sources, which were imaged using natural weighting in order to achieve the best possible 
signal-to-noise. In general we obtain image noise levels close to the theoretical value, usually about 3$\mu$Jy/beam for 6 
hours of on-source time.

\subsubsection{e-MERLIN observations}

Two of the objects (HS~0810+2554 and RX~J0911+0551) were also observed with the e-MERLIN array (Table 2). The 
observations were
carried out at L-band, with a bandwidth covering the wavelength range 1287-1799~MHz. In addition to the target sources, 
observations of nearby phase calibrators were carried out, with a cycle of 7 minutes on the target and 3 on the phase 
calibrator. Additional observations of 3C286 were carried out in order to set the flux scale, and of the bright point
source OQ208 in order to calibrate the bandpass. Data reduction followed standard procedures (Argo 2015) including fringe
fitting to all calibrator sources to determine delays, phase and amplitude calibration using the nearby phase calibrator,
and determination of the flux scale and bandpass calibration, with allowance for the spectral index of the calibrator. 
The telescope weights were modified using standard values for L-band provided by the observatory, and final maps were
made in the AIPS software package distributed by NRAO. Mapping in the case of HS~0810+2554 was complicated by the presence
of a 200-mJy confusing source 6$^{\prime}$ from the target; this source was mapped simultaneously with the target, and
was also used to refine the phase calibration of these observations. Noise levels achieved in these observations were
about 15-30$\mu$Jy, depending on the details of the mapping strategy.

\section{Results and models}

All four objects were detected in these observations, of which all but SDSS~J0924+0219 have individually measured flux 
densities for each lensed image. We discuss the results for each object separately, before making more general remarks
about the measurements. The radio flux densities are given in Table \ref{fluxtable}.

\begin{table}
\begin{tabular}{cccc}
Object & Type & Cpt. & Flux density \\
      &      &      &  (radio,$\mu$Jy)   \\
HE~0435$-$1223 & Cross & A & {\bf 36.0$\pm$2.1}\\
              &       & B & {\bf 26.4$\pm$2.1}   \\
              &       & C & {\bf 34.3$\pm$2.1}  \\
              &       & D & {\bf 16.1$\pm$2.1}   \\
HS~0810+2554   & Fold  & A & {\bf 85.1$\pm$3.7}  \\
              &       & B & {\bf 83.7$\pm$3.7}  \\
              &       & C & {\bf 60.0$\pm$3.7}  \\
              &       & D & {\bf 49.1$\pm$3.7}  \\
RX~J0911+0551  & Cusp  & A & {\bf 26.9$\pm$2.2} \\
              &       & B & {\bf 53.2$\pm$2.2}  \\
              &       & C & {\bf 19.7$\pm$2.2}  \\
              &       & D & {\bf  9.4$\pm$3.0}  \\
              &       & G & {\bf 18.3$\pm$2.2} \\
\end{tabular}
\caption{Radio flux measurements for the sample of four-image lens systems observed with the VLA/e-MERLIN, in which
fluxes can be measured. For HS~0810+2554
the flux densities are from the VLA map at 8.4~GHz; the corresponding flux densities in the e-MERLIN image are (161,173,129,216)
for A,B,C and D respectively, with errors of approximately 30$\mu$Jy in each case.}
\label{fluxtable}
\end{table}

The approach to modelling the observations is the same in each case. First, we make a preliminary assessment of whether
the radio map is consistent with lensing of a point source into point images. We do this by modelling the structure, in
each case, with four point-spread functions (PSFs; Table \ref{fluxtable}), whose extent is known accurately from the 
radio CLEAN procedure. In this model, we fix the separation between the four individual components using 
measurements from archive HST images as reported by the CASTLES astrometry (which are accurate to a few milliarcseconds), 
but the overall registration of the image has been allowed to vary. There are thus six free parameters in the model, 
two for the registration, and four from the flux densities of the individual points. Second, we make a lens model using
constraints from the image-plane radio map; 
for this, we assume a singular isothermal mass distribution (except in the case of HE~0435$-$1223) for the lenses together with a contribution from external
shear. The source is assumed to be of Gaussian profile, and the resulting image plane is compared to the data, optimising
the lens galaxy parameters together with the source position, flux density, size and ellipticity. We note that modern
wide-bandwidth interferometers at centimetre wavelengths, such as the VLA and e-MERLIN, come close to filling the $u-v$ 
plane. Because of this virtually filled aperture, there is no need to fit the data directly in the $u-v$ plane. This
contrasts with the situation in early Atacama Large Millimetre Array (ALMA) datasets used to map sub-millimetre lenses (e.g. Hezaveh et al. 2013).

\subsection{HS~0810+2554}

HS~0810+2554 was discovered by Reimers et al. (2002) and consists of four images with the two southwestern, bright images in 
a merging pair configuration. In HST imaging (Reimers et al. 2002) the lensing galaxy is detected, with an unknown redshift 
(it is estimated as 0.89 by Mosquera \& Kochanek 2011 from the separation and the redshift distribution of existing lenses). 
%
%
These images also show a 0.7-magnitude difference in brightness between the components of the merging pair, contrary to 
the expectations of simple models, but this is likely to be due to microlensing in the lens galaxy. The source is a narrow
absorption line quasar, with relativistic outflows detected using X-ray absorption spectra (Chartas et al. 2014). 
These high velocity outflows may be magnetically driven.

\begin{figure*}
\begin{tabular}{cc}
\includegraphics[width=9cm]{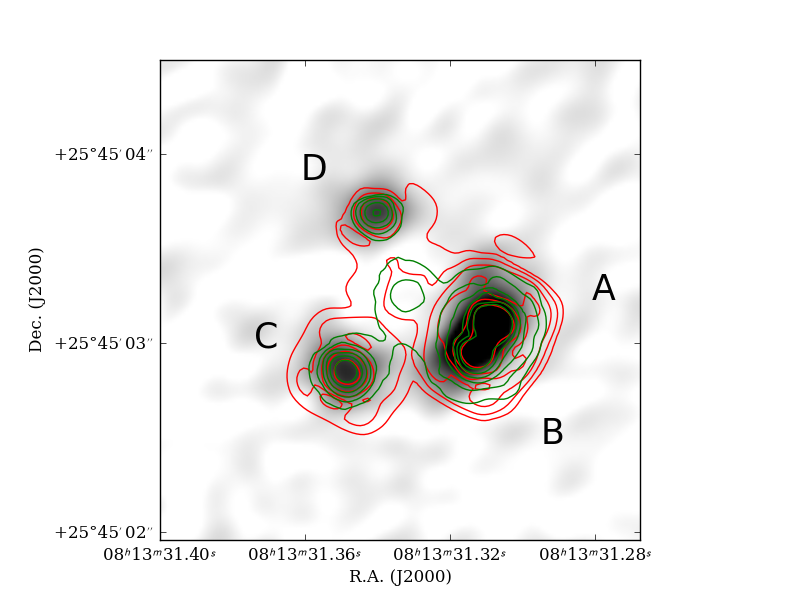}&
\includegraphics[width=7.5cm]{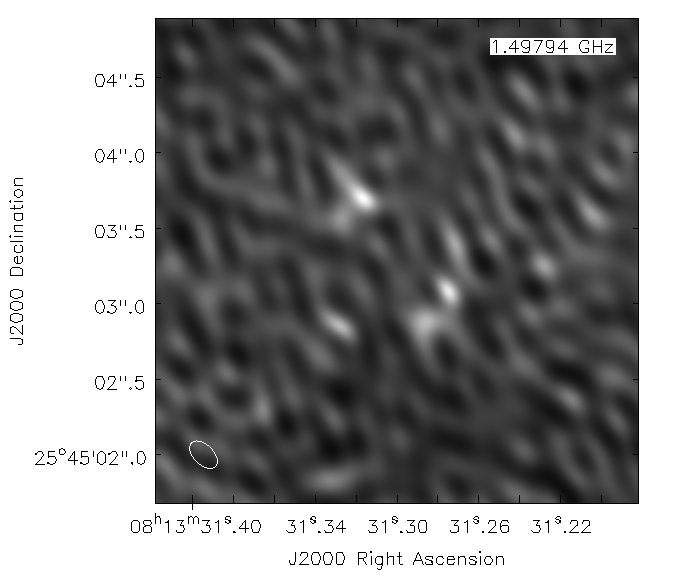}\\
\end{tabular}
\caption{Left: VLA greyscale radio map of HS~0810+2554 at 8.4~GHz. The beam is of full width at half maximum (FWHM) 300$\times$240~mas in position angle
$-65.17^{\circ}$. Archival HST Near-Infrared Camera and Multi-Object Spectrometer (NICMOS, red) and Advanced Camera for Surveys (ACS, green) contours have been performed using image C. The conventional
nomenclature of the images (Reimers et al. 2002) is that the merging complex in the southwest consists of images A and
B, with A being further north. Right: e-MERLIN image of HS~0810+2554 at approximately the same resolution, but a frequency
of 1.6~GHz. The noise level is approximately 29$\mu$Jy/beam; all the images are detected at $>5\sigma$ significance.}
\label{fig_0810}
\end{figure*}

\begin{figure*}
\includegraphics[width=15cm]{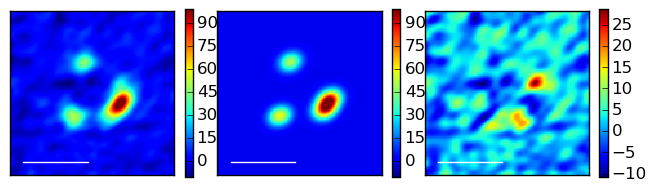}\\
\caption{Model of HS0810+2554, using a point-source model fit (see text). The data, model,  and residual are 
plotted; unlike the case with the extended source model, there appear to be significant features in the residual. 
Note that in this case, and unlike the case of the extended source model, the overall shape of the A-B complex is 
not well reproduced. The bar in each panel represents 1$^{\prime\prime}$, and the colour scales in the sidebars
are in units of $\mu$Jy/beam.}
\label{0810_gauss}
\end{figure*}

Our radio maps from the VLA and e-MERLIN are shown in Fig. \ref{fig_0810}. The components in the VLA image appear 
extended, and a faint arc is visible around the bright 
merging pair. This extension can be quantified by attempting to model the lensed structure only with point sources, with 
separations fixed to that of archival HST data. The best fit shows significant residuals, in particular around the bright
merging components, but also at a lower level around the line connecting images B and C (Fig. \ref{0810_gauss}). It is obvious
from visual inspection that the shape of the A-B complex in the data is significantly different from that of a two-Gaussian
realisation. We therefore conclude that the source is likely to be extended and model it as such.

\begin{figure*}
\includegraphics[width=19cm]{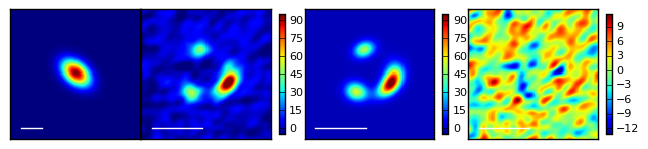}
\caption{Models of HS0810+2554, using a singular isothermal sphere model plus external shear (see text) together with a best-fit extended source. The 
reconstructed source, data, model,  and residual are plotted. The white bar represents 10~mas in the panel of the
reconstructed source, and 1$^{\prime\prime}$ in all other cases. In these and subsequent figures, the data and model
are plotted on the same colour scale, and the residuals are scaled to the minimum/maximum of the residual map. 
The colour-bars represent flux densities in $\mu$Jy/beam; the colour scale of the source is arbitrary, but its
parameters are given in Table \ref{models}.
A good fit is obtained with an unlensed source
size of between 10-15~mas.}
\label{0810_fit}
\end{figure*}

\begin{figure}
\includegraphics[width=10cm]{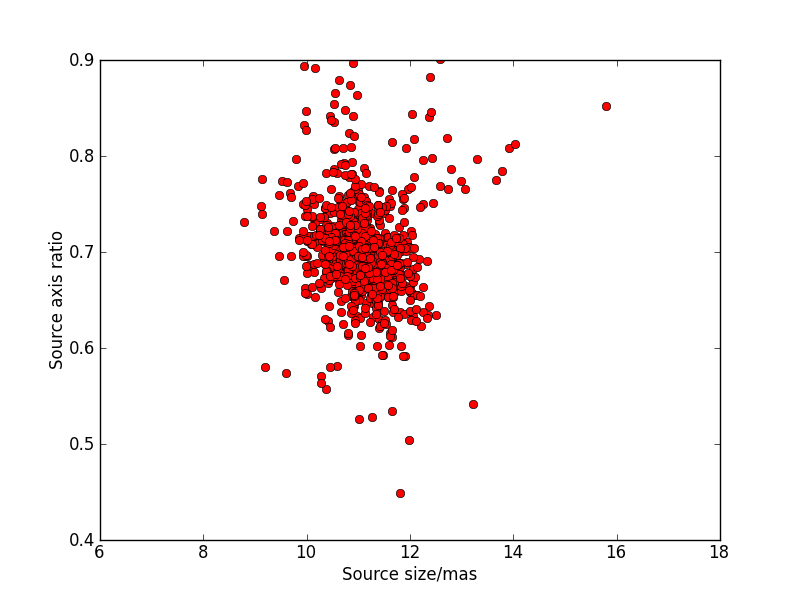}
\caption{MCMC realisations of the model of HS~0810+2554, plotted as probability density of source size
against source axis ratio. The preferred source size corresponds to a linear scale of approximately 100~pc, a unique
resolution for such a faint radio source.}
\label{0810_src}
\end{figure}

In order to model the extended source, we have assumed a simple lens model (singular isothermal ellipsoid plus external 
shear) and treated the source as an ellipse with a flux density, axial ratio, position and orientation which are free to
vary. For each iteration of the model, the source is projected through the lens model, and the result convolved with the 
PSF of the radio map. Minimisation of the $\chi^2$ between the model image and data is carried out, where the $\chi^2$
value is determined from a comparison of the model with the image in all regions where either model or image contains flux 
above 2$\sigma$. Correlations between pixels are neglected. The position of the galaxy is fixed by the use of the Hubble Space Telescope (HST) image during this process. In practice, the
quality of the fit does not depend significantly on this quantity, provided that the source is allowed to move to keep the
same distance between it and the galaxy. An acceptable fit of $\chi^2$=1.6 is obtained with such a procedure; the parameters 
of this fit are given in Table \ref{models} and shown in Fig. \ref{0810_fit} and \ref{0810_src}. The two close images, A and B, are of
approximately the same flux density, as expected in the absence of millilensing-scale substructure and in contrast to 
the measurements in the optical and near infrared. The model implies a magnification of about 25 for the brightest image,
yielding an intrinsic flux density of 3.5~$\mu$Jy for the source. The implied magnification is a factor of 2 less than that 
of the model by Assef et al. (2011), but HS~0810+2554 is in the high-magnification regime where the source is contained 
within a very small astroid caustic, and minor movements in the source position produce major changes in the implied
magnification. 

\begin{table*}
\begin{tabular}{cccc}
Quantity               & HS0810+2554                           & HE0435$-$1223         & RXJ0911+0551\\ \hline
Source position/mas    & 0.1~E, 13.0~S                         & -68~E, 18~S           & 468~E, 28~S\\
Source FWHM along major axis/mas  & 12$\pm$1                              & 80$^{+5}_{-5}$        & 131$^{+15}_{-11}$\\
Source flux/$\mu$Jy    & 3.6$\pm$0.2                           & 2.9$^{+0.3}_{-0.4}$   & 3.7$^{+0.3}_{-0.2}$\\
Source $b/a$           & 0.66$^{+0.06}_{-0.09}$                & $\equiv$1.0           & $\equiv$1.0\\
Source position angle  & (47$\pm$5)$^{\circ}$                  & --                    & -- \\
Galaxy critical radius/mas & 473$\pm$10                        & 1138$^{+19}_{-6}$     & 1047$^{+11}_{-38}$\\
Density slope (2 = isothermal)          & $\equiv$2.0                           & 2.00$^{0.08}_{-0.06}$ & $\equiv$2.0\\
Galaxy ellipticity     & 0.0003$\pm$0.0003                     & 0.26$\pm$0.02         & 0.15$^{+0.03}_{-0.09}$\\
External shear         & 0.023$\pm$0.006, (29$\pm$4)$^{\circ}$ & 0.039$^{+0.004}_{-0.011}$, ($-$30$\pm$7)$^{\circ}$&0.373$^{+0.033}_{-0.011}$, (9$\pm$2)$^{\circ}$
\end{tabular}
\caption{Model fitting results for the three lens systems for which lens modelling is possible (all observed
lenses except SDSS~0924+0219). The source position is quoted relative to the galaxy position, and all distances are given in units of milliarcseconds. For HE0435$-$1223 and RXJ0911+0551, the galaxy critical radius corresponds to the Einstein radius measured along the major axis.}
\label{models}
\end{table*}


In order to derive uncertainties, the parameters have been run through a Markov Chain Monte Carlo (MCMC) process 
using the publicly available {\sc emcee} routine (Foreman-Mackey et al. 2013). We have assumed a number of hard
priors, namely limits of 0$^{\prime\prime}$--1$^{\prime\prime}$ for the Einstein radius of the galaxy, limits of
$0<\gamma<0.5$ for the external shear contribution, a requirement that the source flux and source width are positive,
and that ellipticities of the source and galaxy must be $<$1. Finally, we have imposed a Gaussian prior on the 
position of the lens galaxy, based on the measured position in the HST image and with a width of 10~mas. The 
results (Table \ref{models}) make it clear that the source is extended by approximately 10~mas in the source plane, 
corresponding to about 70~pc in physical scale, at a position angle of approximately 50$^{\circ}$. As usual with
strong lens systems, we obtain a relatively accurate measurement of the Einstein radius of the lens galaxy, which
is modelled as being almost circular. This is consistent with its appearance on archival HST images.

Although the existing e-MERLIN images do not have very high signal-to-noise, they do allow us to measure an approximate
overall spectral index, because the resolution of the e-MERLIN 1.6-GHz image is very similar to that of the VLA at
8.4~GHz. This spectral index is moderately steep, at $-0.55\pm0.1$.

\subsection{HE~0435$-$1223}

HE~0435$-$1223 was discovered by Wisotzki et al. (2002) and identified as a four-image system with an early-type lens galaxy. 
The lens redshift was obtained by Morgan et al. (2005), who also found that the lens is part of a group of galaxies. 
Microlensing was detected in a subsequent monitoring campaign (Kochanek et al. 2006) which probably affects the A component 
most strongly (Ricci et al. 2011, Courbin et al. 2011), and it has also been shown that the broad-line region in this 
object is subject to microlensing (Sluse
et al. 2012; Braibant et al. 2014). Modelling of the lens is able to reproduce well the positions of the lensed images, using
only a single-galaxy deflector model (Sluse et al. 2012). Fadely \& Keeton (2012) examined and modelled this object extensively 
in a search for evidence of substructure in the lensing galaxy, using their $L'$-band mid-infrared fluxes of the four components 
as inputs for the modelling. Since much of the quasar mid-infrared emission comes from a circumnuclear torus, this may be 
immune to microlensing as the torus is likely to be relatively large.

\begin{figure}
\includegraphics[width=10cm]{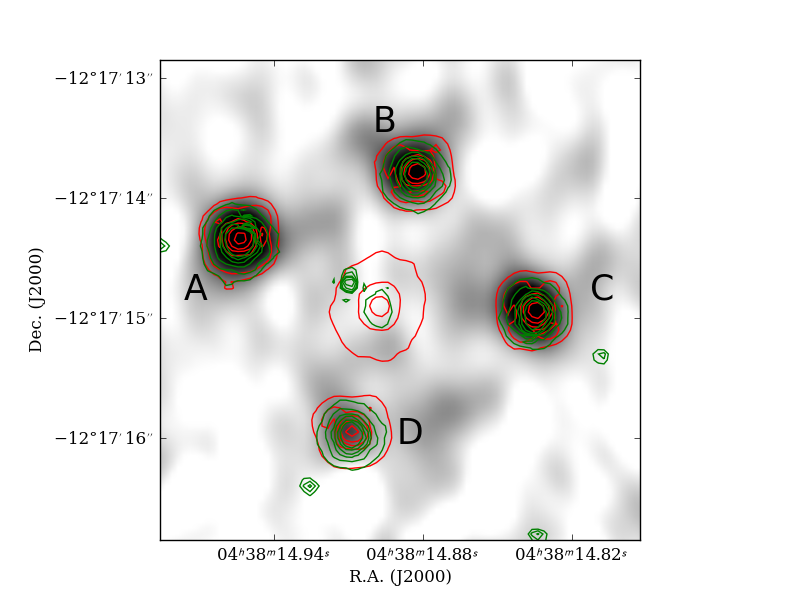}
\caption{VLA radio map of HE~0435$-$1223. The greyscale runs from 0 to 20$\mu$Jy/beam. The beam is circular, and of 
FWHM 0\farcs45. Archival NICMOS (red contours) and ACS (green contours) images are overlaid. Registration of these images 
has been performed using image A. The conventional nomenclature of the images is that A is the easternmost and B,C,D 
proceed clockwise around the lens galaxy.}
\label{fig_0435}
\end{figure}

\begin{figure*}
\includegraphics[width=13cm]{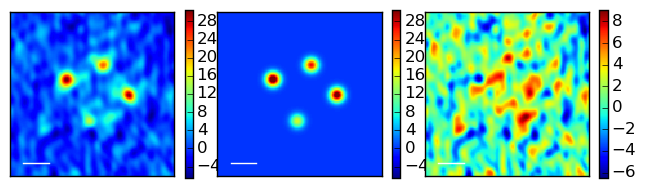}\\
\caption{Models of HE~0435$-$1223, using a point-source model fit (see text). The data, model,  and residual are 
plotted; unlike the case with the extended source model, there appear to be significant features in the residual. 
The bar in each panel represents 1$^{\prime\prime}$. The numbers on each colourbar are in units of $\mu$Jy/beam.}
\label{0435_gauss}
\end{figure*}

\begin{figure*}
\includegraphics[width=16cm]{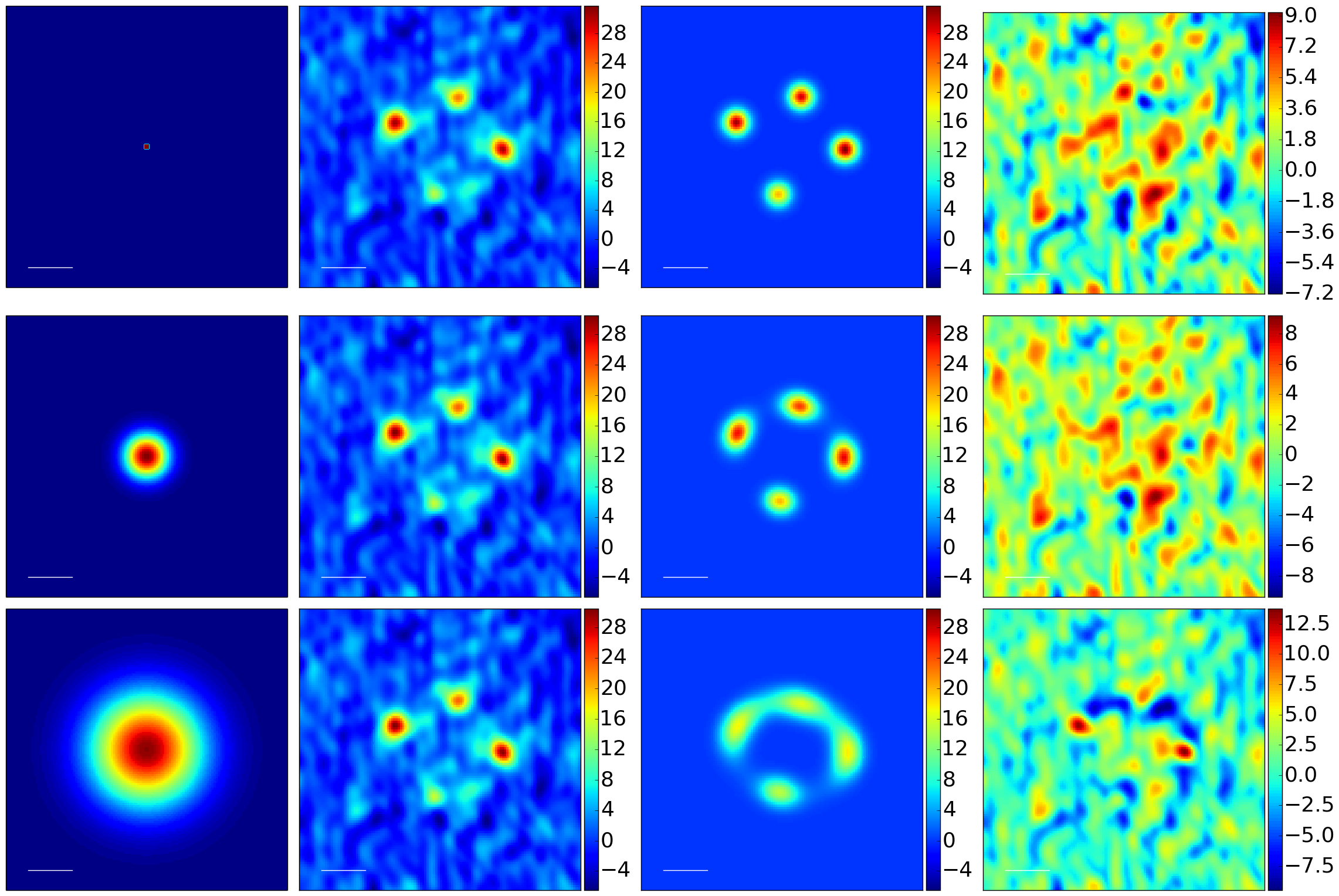}
\caption{Models of HE~0435$-$1223/ From left to right: source model; data; image-plane model; residual. The white
bars represent 100~mas in the source plane (left panel) and 1 arcsecond in all other panels. Three different fits are shown: (top) small 3~mas source, (middle) optimal 80~mas source, (bottom) large 200~mas source. Although the 80 mas source is preferred by our MCMC analysis, smaller sources provide an equally good visual fit, while larger sources lead to significant model residuals. The colour-bars represent flux densities in $\mu$Jy/beam; the colour scale of the source is arbitrary, but its parameters are given in Table~\ref{models}.}
\label{0435_extended}
\end{figure*}


Our radio map (Fig. \ref{fig_0435}) shows all four lensed images, at positions negligibly different from those of the optical 
and infrared HST images (obtained from the HST archive). Once again, therefore, we have modelled the radio map assuming that
it consists of four point sources, whose separation is determined by the HST optical image. The results of this procedure are
shown in Fig. \ref{0435_gauss}. There are hints of emission outside the four point sources, particularly close to image C, but
these are at the 2-$\sigma$ level and would need deeper observations to confirm or rule out. The image fluxes are given in Table 
\ref{fluxtable}, and, with a ratio of 1.05:0.77:1.00:0.47 between A:B:C:D images, differ significantly from the $L^{\prime}$ 
ratios 1.71:0.99:1.00:0.81 of Fadely \& Keeton (2012). In particular, the $A/C$ and $C/D$ ratios differ by about 3$\sigma$ between 
the two sets of observations, the difference in $C/D$ ratio being particularly obvious from Fig. \ref{fig_0435}.

This difference in flux density ratios, and its explanation, is a difficult and intriguing problem. Fadely \& Keeton (2012)
undertook extensive modelling of this system using a softened power-law for the primary galaxy, together with a singular
isothermal model for the nearby galaxy, G22, which is about four arcseconds to the SW. They also included a shear component, to
account for the more general shear field of the cluster. Smooth models with a slightly shallower slope than isothermal failed 
to reproduce the infrared fluxes, and further analysis showed that the Bayesian evidence favoured a contribution due to 
substructure around A. The observation driving this conclusion was the A/C ratio, which smooth models preferred to be between 
1.4 and 1.5, as opposed to the higher value in the infrared data. Our radio data, on the other hand, prefer a much lower value 
for the A/C ratio, together with a much fainter D component. Since the radio is almost certainly not affected by microlensing,
this is a puzzling result.

In our next step of modelling the data, we assume the source is point-like and include the observed time delays (Courbin et al. 2011) as additional constraints.
Modelling the lens as an ellipsoidal power-law with external shear and a SIS at the position of G22 ($z=0.78$, Chen et al. 2014), realistic, smooth models are unable to reproduce the data.
The best model, further constrained using strong Gaussian priors on the positions ($1\sigma$=10~mas) of the galaxies, agrees with those observed time delays within the errors but not with the image fluxes, yielding a $\chi^2$ of 14.9 for four degrees of freedom.\footnote{\label{ftnt:ndof}Because of the intrinsic nonlinearity of the lens modelling and the strong Gaussian priors placed on the galaxy positions, calculating the the number of degrees of freedom is nontrivial. Thus, we have chosen to count each galaxy position parameter as half a degree of freedom, and so the ``true'' number of degrees of freedom may be $\pm 2$ of the number we quote here.}
In particular, the predicted B/C and D/C ratios are in disagreement at the $1.3\sigma$ and $3\sigma$ levels, respectively.
For the main lensing galaxy, the best model prefers an ellipticity of $e\approx 0.28$ and power-law slope of $\gamma^\prime\approx 2.24$.
We note that the steep density slope may be driven by the model trying to fit the flux ratios.

The flux ratio anomaly seen with the smooth model could be explained by invoking substructure.
To explore this possibility, we follow the approach of Fadely \& Keeton (2012).
Briefly, substructure clumps are modelled using a pseudo-Jaffe profile, and a wide range of masses are considered, whose masses enclosed within the Einstein radii range from $10^4-10^9$~$M_\odot$.
Modelling the smooth lens component as before, we find that clumps placed near images A, B, or C do not improve the fit.
However, clumps over a large range of masses placed near image D can bring the model into good agreement with the data, yielding a $\chi^2$ of 1.5 for one degree of freedom.\footnotemark[\getrefnumber{ftnt:ndof}]
Like Fadely \& Keeton (2012), we find that more massive clumps can be placed farther away from the image, while less massive clumps must be placed nearer.
Furthermore, the clumps cannot lie within approximately two Einstein radii of image D.
Otherwise, image splitting can magnify the image, making matters worse.
As for the main lensing galaxy, ellipticities of $e\approx 0.33$ and steeper-than-isothermal power-law slopes of $\gamma^\prime\approx 2.33$ are preferred.

Alternatively, if we do not invoke substructure, another possible solution arises if the radio emission region is extended and differentially magnified (see e.g. Serjeant 2013).
Because the size of the caustic is less than approximately 400~mas, a wide range of source sizes below this scale can reproduce the data.
We again use a two-deflector model, including the main lensing galaxy as an ellipsoidal power-law with a contribution from external shear and G22 as a SIS.
The source is modelled as a spherical Gaussian.
Owing to the large number of image pixels (3150 pixels) used to constrain the model, including time delay constraints for a point source at the position of the source does not significantly affect our results, and so we include them for consistency with the previous analyses.
Additionally, to try to account for the noise correlation in the data and to be conservative in our parameter inference, we follow the suggestion of Riechers et. al. (2008) and increase the input noise level by a factor dependent on the noise correlation length scale (a factor of three, in this case).
Our best model achieves a $\chi^2$ of 854 for 3141 degrees of freedom.\footnotemark[\getrefnumber{ftnt:ndof}]
After marginalizing over all lens model parameters, our MCMC analysis finds that an isothermal slope is preferred for the main lensing galaxy ($\gamma^\prime = 2.00^{+0.08}_{-0.06}$) and that the source is of FWHM 80$^{+5}_{-5}$~mas.
This optimal source size leads to an image-plane model that shows discernible extended structure (Fig.~\ref{0435_extended}).
By visual inspection, we find that source sizes more than an order of magnitude smaller\footnote{The image plane is appropriately oversampled to ensure that fluxes are calculated accurately.} can also fit the data reasonably well but result in point-like images at these resolutions and leave larger model residuals.
Sources a factor of two larger, on the other hand, are clearly unfavorable by both visual inspection and the MCMC analysis.

Of the possible choices for explaining the data, we prefer the option that the source is extended and differentially magnified.
As radio sources are likely to be more extended than their optical or mid-infrared counterparts, this seems to be the most natural choice.
Furthermore, finite source size effects would likely be required in order to simultaneously explain the flux ratio anomalies in the mid-IR as well.
We note, however, that due to the non-Gaussianity and correlation of the noise in the immediate regions surrounding the lensed images, we do not strictly limit our conclusions by the results of the MCMC analysis.
Instead, we provide a conservative upper limit of 200~mas for the source size.
For sources larger than this, a visual inspection of the model residuals clearly shows that the source has become too large.

\subsection{RX~J0911+0551}

\begin{figure}
\includegraphics[width=10cm]{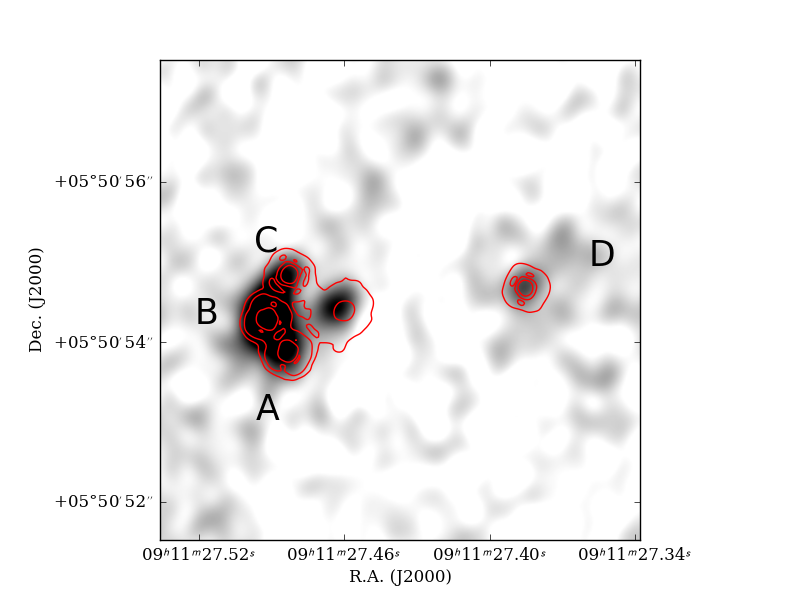}
\caption{JVLA radio map of RX~J0911+0551 (greyscaled from 0 to 20$\mu$Jy) with contours from archival HST/NICMOS data superimposed, aligned on image D. 
The beam is of FWHM 560$\times$390~mas in position angle $-39.2^{\circ}$.
The three close images to the left are A,B,C (from south to north), and image D is at the right of the picture. The lensing galaxy 
(between the images) is radio-loud, with a flux density of about 16$\mu$Jy.}
\label{fig_0911}
\end{figure}

RX~J0911+0551 (Bade et al. 1997) is a cusp-configuration lens system, with three close images (A,B and C) and a fourth image
some distance to the west. The mass environment is relatively complicated; the lens lies close to a massive cluster about 
40$^{\prime\prime}$ away and in addition to the primary galaxy, a second galaxy lies close to the system, complicating the process
of mass modelling. Our VLA 5-GHz image is shown in Fig. \ref{fig_0911} and has a noise level of just under 2$\mu$Jy/beam. All four
lensed images are clearly detected in the radio map, and in addition we detect radio emission at the position of the lensing galaxy. 
We do not detect any of the components in the e-MERLIN 1.5 GHz image, which has a noise level of 16$\mu$Jy/beam. 




To quantify the non-detection, and thus derive limits on spectral index between the e-MERLIN and VLA observations, the e-MERLIN
observation was conservatively simulated with four components of the size of the VLA beam ($\sim$500~mas), whose flux densities
were given by the VLA map. Gaussian noise was added to the map at the same level as the observations (i.e. RMS 16$\mu$Jy). The
components were used to generate a $u-v$ dataset with the sampling function and noise level of the actual e-MERLIN observations.
This was imaged and the lower limit on the spectral index resulted from the input fluxes for which the components could just not 
be recovered from the simulated image by model fitting.  The lower limit on the spectral index was found to be $\alpha=-0.5$.

\begin{figure*}
\begin{tabular}{c}
\includegraphics[width=15cm]{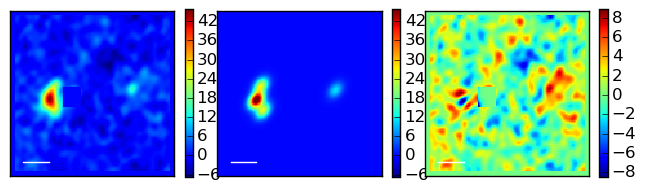}\\
\includegraphics[width=19cm]{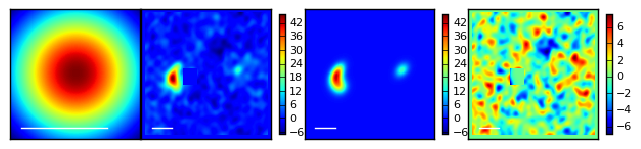}\\
\end{tabular}
\caption{Models of RXJ~0911+0551, using a point-source model fit (above, see text). The data, model,  and residual are 
plotted. In this case, the point-image model appears to have significant difficulty in fitting the A-B-C complex at the eastern
end of the system. The bar in each panel represents 1$^{\prime\prime}$. Model using an extended source (below). The panels contain
the source plane, the data, the model and the residual. The area around the galaxy has been blanked and excluded from the fit. In all cases the numbers on the colourbars are in $\mu$Jy/beam; the source panel
colourscale is arbitrary, but the source parameters are given in Table \ref{models}.}
\label{0911_gauss}
\end{figure*}

We have again attempted to model the image plane, represented by the VLA map, as a sum of point spread functions whose relative
position is determined by the HST astrometry (Fig. \ref{0911_gauss}). Here it does appear that point models have difficulty in
reproducing the structure, in particular the shape of the A-B-C complex, although we recover good estimates for the flux 
densities of each image (Table \ref{fluxtable}).
The flux densities of the A,B and C images in the radio have a ratio very close to 1:2:1, close to that
expected by the cusp relation and suggesting that substructure does not need to be invoked in this case. This contrasts with the
optical flux densities, and in particular with the flux density ratio 2:2:1 between A, B and C measured by Burud et al. (1998).
Hence the optical measurements are almost certainly affected by microlensing. By contrast, Sluse et al. (2012) model this lens 
using astrometric constraints and a model consisting of a singular isothermal ellipsoid plus shear. They obtain image flux ratios 
(A:B:C:D) of 1:1.87:0.88:0.34. Our corresponding values are 1:2.05:0.73:0.35, consistent within the errors with 
Sluse et al.'s model. The source flux density predicted by this lens model is about 5$\mu$Jy.

Motivated by the residuals observed in the point-source model, we have again fitted a model in which a Gaussian-shaped extended
source is lensed. Provided that the source size is not very small, neither it nor the source shape is well constrained (Table 
\ref{models}). The modelled shear is large, suggesting that we are seeing the influence of the cluster to the south. The good
overall fit to the data, $\chi^2$=3518 for 2290 degrees of freedom, gives no significant evidence for effects of substructure in the lens galaxy.

\subsection{SDSS~J0924+0219}

SDSS~J0924+0219 (Inada et al. 2003b) is a very interesting object because it has such a large optical flux anomaly, almost 
certainly as a result of microlensing (Morgan et al. 2008). This anomaly decreases with increasing observation wavelength (Floyd, 
Bate \& Webster 2009) allowing the size of the accretion disk to be measured, but the anomaly persists over a period of $\sim$7~years
(Faure et al. 2011) raising the possibility that some of it may be due to the effects of lensing by substructure.

\begin{figure}
\includegraphics[width=10cm]{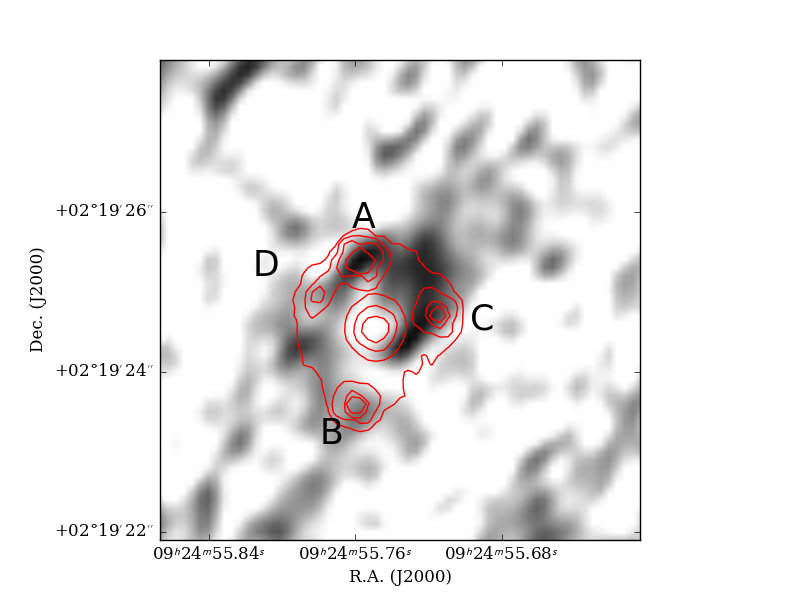}
\caption{VLA radio map of SDSS~J0924+0219, with greyscale from
0 to 20$\mu$Jy/beam, and a restoring beam of 696$\times$416~mas in PA~$-43^{\circ}$. The object is detected,
but individual flux densities for the images are impossible to measure.
Image registration to archival HST/NICMOS data (contours) has been done
by eye, but this procedure is not well-defined to better than the
absolute astrometry of the HST image.}
\label{fig_0924}
\end{figure}

Although we detect the object at a reasonable level of significance (Fig. \ref{fig_0924}), we unfortunately do not have sufficient 
signal-to-noise, in the three hours of observation time allocated to this object, to measure the flux densities separately. At a 
total flux density of $\sim 15\mu$Jy, this is by far the weakest of the objects studied. Using an isothermal model, plus external
shear, fitted to the positions reported by Inada et al. (2003b), we obtain magnifications for the four components (A, B, C and D)
of approximately 13, 5, 5 and 11. Our overall flux density of $\sim 50\,\mu$Jy implies an unlensed source flux density of about
1.5$\mu$Jy. This is the second faintest radio source yet detected, the faintest being SDSS~J1004+4112 (Jackson 2011); further 
observations of the sample of radio-quiet quasars are likely to yield the first detected nano-Jy radio source.

\section{Discussion and conclusions}

\subsection{Radio properties of the lensing galaxies}

We detect the lensing galaxy in only one of the four objects: RX~J0911+0551, at a level of 18$\mu$Jy. The radio flux 
density of the lensing galaxies in the other three objects is $<6-8\mu$Jy (3$\sigma$).

The lensing galaxy flux density in RX~J0911+0551 corresponds to a luminosity of $5\times 10^{22}$~W$\,$Hz$^{-1}$, which is
at the top of the range that can plausibly be attributed to star formation; this range is bounded by the $10^{21}$~W$\,$Hz$^{-1}$
of the Milky Way and 10$^{23}$~W$\,$Hz$^{-1}$ for extreme star-forming galaxies. It is also close to the boundary between
star-forming radio emitters and AGNs found by Kimball et al. (2011) in their study of radio luminosity functions of 
nearby ($z\sim 0.2$) AGN. There is no evidence for ongoing star formation in the lens galaxy, which seems to be an early-type
galaxy, in other wavebands. For example, Burud et al. (1998) conducted optical and near-IR imaging and found that the lensing 
galaxy has a similar red colour to that of nearby members of the cluster of which it is part.

\subsection{The nature of radio-quiet quasars}

This work has resulted in the measurement of radio flux densities, and in some cases resolved radio structure, in a significant 
number of radio-``quiet'' lensed quasar systems. Such radio sources, if unlensed, would be beyond the reach of current instruments 
in all but exceptional observing times, and are objects whose study will only become routine with the SKA. The luminosity
of an object of intrinsic flux density of 1$\mu$Jy and flat spectral index is about 1.0$\times 10^{21}$WHz$^{-1}$ at $z=0.5$, 
5$\times 10^{21}$WHz$^{-1}$ at $z=1$ and 30$\times 10^{21}$WHz$^{-1}$ at $z=2$, orders of magnitude below what is typically 
accessible with current surveys except at low redshift (e.g. fig. 4 of Condon et al. 2013). Previous studies of radio-quiet
quasars have focused on optically bright quasars, such as the Palomar Green sample (Kellermann et al. 1994). In accordance with
the radio-optical correlation noticed by White et al. (2007), these objects have typical radio flux densities of a few hundred 
$\mu$Jy, two orders of magnitude brighter than the intrinsic flux densities of the objects studied here. 

We have observed four of the 15 known optically-selected, four-image quasar lenses with $\delta>-20^{\circ}$, and all of them have 
intrinsic radio flux density of between 1 and 5$\mu$Jy. Of the other nine, three are known to have significant radio
emission. PG1115+080 has VLA archival data at 8.4~GHz taken in the compact (D) configuration, which yield a total radio flux density
of 153$\pm$17~$\mu$Jy, although the resolution of a few arcseconds does not allow the flux density of individual components
to be determined. However, the likely magnifications in this lens system suggest that the intrinsic flux density of the source
is also a few $\mu$Jy. A similar result can be derived for the lens system RX~J1131$-$1321, which was found to have significant
radio emission by Wucknitz \& Volino (2008). Finally, H1413+117 is a radio-intermediate object which has been studied with
the VLA by Kayser et al. (1990). Further lenses from the COSMOS survey (Faure et al. 2008, Jackson 2008) do not have significant
radio emission (Schinnerer et al. 2007, Schinnerer et al. 2010) in the VLA-COSMOS survey, and three quad lens systems from the
Sloan Quasar Lens Search (SDSS~J1138+0314, SDSS~J1251+2935, SDSS~J1330+1810) do not yet have deep radio imaging. It is therefore
likely that at least half of optically-selected quasar lens systems will show radio emission at the micro-Jansky level, if
examined carefully, and a more complete census will be the focus of future work.

\begin{figure}
\includegraphics[width=8cm]{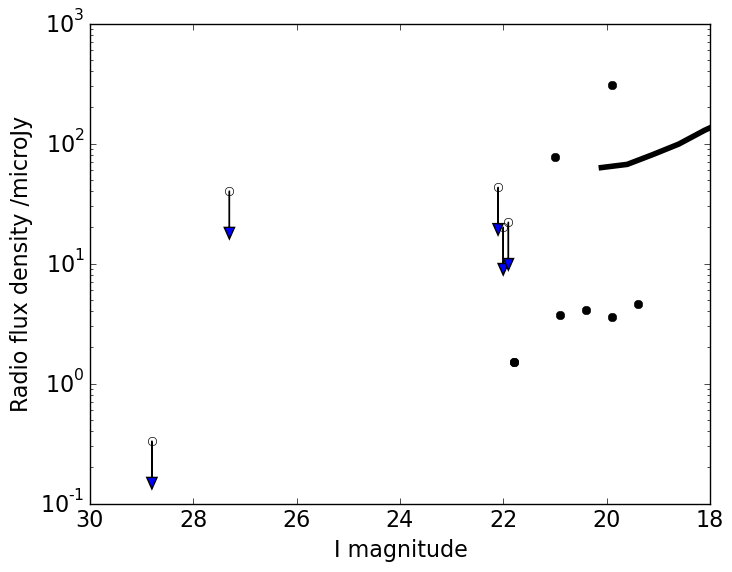}
\caption{Intrinsic radio flux densities versus intrinsic I-band magnitudes for a sample of optically-selected quasar lenses.
Both radio and optical flux densities have been demagnified using lens models. 
Data and models have been derived from Ratnatunga, Griffiths \& Ostrander 1999, Wisotzki et al. 2002, Reimers et al. 2002,
Burud et al. 1998, Inada et al. 2003 a,b, Ghosh \& Narasimha 2009, Anguita et al. 2009, Jackson 2011, Wucknitz \& Volino 2008,
Kayo et al. 2007, Oguri et al. 2008, Assef et al. 2011, in some case supplemented by further modelling. The locus of radio vs. optical flux densities reported by White et al. (2007) is sketched near the top right.}
\label{radopt}
\end{figure}

The existing data are plotted in Fig. \ref{radopt}. The radio fluxes have been derived from the literature (see the figure
caption), with in some cases a limit of 1~mJy inferred from their absence from the FIRST 1.4-GHz catalogue. The current sample
is small. However, we note that the median radio flux density inferred by White et al. (2007) from their stacking analysis is
about 50-70~$\mu$Jy at $20<I<21$. This, combined with the distribution of detections in our optically-selected sample, suggests
that there is a large scatter in radio flux densities at this optical magnitude, if not an outright bimodality.

There are a number of theoretical models for the origin of radio emission in radio quiet quasars, each of which makes rather
different predictions for what should be observed. The first possibility is that of a smaller version of a radio-loud quasar,
where a flat-spectrum radio core and steeper-spectrum radio jet emission are present in some ratio (Urry \& Padovani 1995,
Ulvestad et al. 2005). In this case, we would expect steep-spectrum emission on scales of parsecs to tens of parsecs in 
addition to a compact, flat-spectrum radio core. A second possibility is the emission from radio starbursts in 
a similar manner to nearby examples such as M82 (e.g. Condon et al. 2013). This emission is expected to be optically
thin synchrotron from supernova remnants, but should extend over most of the galaxy disk and therefore have a 
characteristic size of about 1 arcsecond, or 5-10~kpc; studies of star-forming radio sources in the HDF, with mean redshifts
around 1, have shown that the radio emission nearly always displays a similar characteristic size (Muxlow et al. 2005). 
Alternatively, the radio emission could be produced by thermal processes 
close to the accretion disk. Suggestions for this include optically thin free-free emission from a disk wind (Blundell \&
Kuncic 2007) and emission from magnetically-heated coronae (Laor \& Behar 2008). In both cases the emission would be expected
to originate close to the centre. In the case of disk winds, this scale is likely to be at least 0.1--1pc, but for magnetically
heated coronae the scale would be smaller. In both these cases, however, the radio spectral index should be approximately flat.

We can use our data to confront the models in a number of ways. The first relevant result is the source sizes implied by our
lens modelling, which are of the order of 70~pc in HS~0810+2554. In HE~0435$-$1223 we find a characteristic size which is
more uncertain, but probably greater than a few milliarcseconds and certainly less than 200~mas ($<$2~kpc). In RX~J0911+0551
we again detect a significantly extended source, but whose size is likely to be 100-150mas ($\sim$1~kpc) rather than the larger
values which would be expected for a typical star-forming region in a radio source. Either we have an unrepresentative sample,
or the star-formation model is disfavoured compared to the non-thermal AGN hypothesis. This is in contrast to the inference
of star-formation as the cause of the radio emission, made by Wucknitz \& Volino in the case of RX~J1131$-$1231. A definitive 
test should be available using VLBI, as a non-thermal source should contain high-brightness emission at the $\mu$Jy level which 
is detectable with current VLBI sensitivities.

A second, although rather more equivocal, result concerns the measured spectral indices. The spectral index of the radio
emission in HS~0810+2554 appears to be steep, consistent with synchrotron emission from either a non-thermal source or a
star-forming component. Both the spectral index and the characteristic size disfavour coronal emission and emission from
disk winds, which would be expected to be relatively flat-spectrum and to be emitted from a smaller region. However, the VLA and e-MERLIN observations of RX~J0911+0551 may be consistent with such models. Because the spectral index limit is currently relatively
loose ($\alpha>-0.5$), further e-MERLIN observations are needed to make a more definite statement.



\subsection{Substructure in lensing galaxies}

Radio flux densities in four-image lens systems are important because they potentially give indications of substructure in
lensing galaxies (or along the line of sight), in the form of flux anomalies (Mao \& Schneider 1998; Dalal \& Kochanek 2002). 
In particular, violations of the cusp and fold relation allow us to quantify the levels of substructure present (e.g.
Xu et al. 2009, Xu et al. 2015). Flux ratios between images can also be affected by propagation effects (scattering in the
case of the radio waveband), microlensing (in the case when the source is smaller than the intrinsic size of the Einstein
radius of stars in the lensing galaxy, or about 1~$\mu$arcsec), variability (in the case of a source which varies significantly
over the time delay between the images) and source size (which can affect flux ratios, e.g. Amara et al. 2006; Metcalf \& Amara
2012). The use of cusp and fold relations, rather than the observation of disagreement with a smooth model, is important because
otherwise the effect of substructure on the image flux ratios can be partially absorbed by varying the smooth model.

We detect no new evidence in our objects for substructure. Indeed, the flux ratios of RX~J0911+0551 obey the cusp relation
within the errors of the measured radio flux densities. The flux ratios of the fold system HS~0810+2554 are also consistent
with a smooth model, as the brightnesses of the merging images are equal to within the errors, unlike the optical case in
which microlensing affects one of the images. The case of HE~0435$-$1223 is more interesting. Here we obtain flux density ratios
which are inconsistent with not only the optical, but also the mid-IR bands. There are a number of explanations for
this which we do not favour. Radio microlensing could affect the radio flux densities, but would require a very small radio
source size (microarcseconds rather than milliarcseconds). Variability of the radio source, together with a time-delay, is
also possible. However, intrinsic variations in typical radio-faint quasars are on timescales of several weeks to months
with fluctuations at the $10-20$\% level (Barvainis et al. 2005). In order to reproduce these observations, a variation in
the source flux of $\sim$40\% would be required within the time-delay scales measured by Courbin et al. (2011) ($-$6.5 and
$-$14.3 days for B-D and C-D respectively). Flux anomalies due to variations in the mid-IR are still less likely, because
the $L'$-band is expected to have a significant contribution from a dusty torus, whose size is $\gtrsim 1\textnormal{pc}$,
implying a light crossing-time of several years or more. Differential extinction due to dust at non-radio wavelengths is 
not a likely culprit, as the colours (Wisotzki et al. 2002) and the continuum slopes (Morgan et al. 2005; Wisotzki et al. 2004) 
are nearly identical for the four lensed images. While substructure can be used to explain the various flux ratio anomalies 
seen in the optical, mid-IR, and radio, it may prove difficult and would perhaps require fine-tuning to simultaneously explain 
all the observations with substructure alone. Instead, at least some of the explanation is likely to be the effects of
finite source sizes; especially given the results in the other objects, an intrinsic radio source size of order a few 
parsecs is the explanation that we favour.

\section*{Acknowledgements}

The Karl G. Jansky Very Large Array is operated by the U.S. National Radio Astronomy Observatory. NRAO is a facility of the U.S. National Science Foundation operated under cooperative agreement by Associated Universities Inc. e-MERLIN is operated by the University of Manchester at the Jodrell Bank
Observatory on behalf of the UK Science and Technology Facilities Council. We thank Ian Browne and an anonymous referee for comments on
the paper.

\section*{References}

\small
\parindent 0mm 

Amara, A., Metcalf, R.B., Cox, T.J., Ostriker, J.P., 2006, MNRAS, 367, 1367.

Anguita, T., et al., 2009, A\&A, 507, 35

Argo, M., 2015. Astrophysics Source Code Library, ascl:1407.017.

Assef, R.J., et al., 2011, ApJ, 742, 93

Baars, J.W.M., Genzel, R., Pauliny-Toth, I.I.K., Witzel, A., 1977, A\&A, 61, 99.

Bade, N., Siebert, J., Lopez, S., Voges, W., Reimers, D., 1997, A\&A, 317L, 13.

Bartelmann, M., 2010, CQGra, 27w3001B, .

Barvainis, R., Leh\'ar, J., Birkinshaw, M., Falcke, H., Blundell, K.M., 2005, ApJ, 618, 108.

Becker, R.H., White, R.L., Helfand, D.J., 1995, ApJ, 450, 559.

Belokurov, V., et al., 2006, ApJ, 647, L111.

Biggs, A.D., et al., 2004, MNRAS, 350, 949.

Blackburne, J.A., Pooley, D., Rappaport, S., Schechter, P.L., 2011, ApJ, 729, 34.

Blundell, K.M., Beasley, A.J., 1998, MNRAS, 299, 165.

Blundell, K.M., Kuncic, Z., 2007, ApJ, 668L, 103.

Bradac, M., et al., 2004, A\&A, 423, 797.

Braibant, L. Hutsem\'ekers, D. Sluse, D. Anguita,T., Garcia-Vergara, C.J., 2014, A\&A, 565, L11

Browne, I.W.A., et al., 2003, MNRAS, 341, 13.

Burud, I., et al., 1998, ApJ, 501L, 5.

Burud, I., et al., 2002, A\&A, 383, 71.

Chartas, G. et al. 2014, ApJ, 783, 57

Chen, H.-W., Gauthier, J.R., Sharon, K., Johnson, S.D., Nair, P., Liang, C.J., 2014., MNRAS, 438, 1435.

Chiba, M., 2002, ApJ, 565, 17.

Chiba, M., Minezaki, T., Kashikawa, N., Kataza, H., Inoue, K.T., 2005, ApJ, 627, 53.

Condon, J.J., et al., 1998, AJ, 115, 1693.

Condon, J.J., Kellermann, K.I., Kimball, A.E., Ivezic, Z, Perley, R.A., 2013, ApJ, 768, 37

Congdon, A.B., Keeton, C.R., Nordgren, C.E., 2008, MNRAS, 389, 398.

Courbin, F., Saha, P., Schechter, P.L., 2002, LNP, 608, 1.

Courbin, F., et al., 2011, A\&A, 536, 53

Dalal, N., Kochanek, C.S., 2002, ApJ, 572, 25.

Dye, S., Warren, S.J., 2005, ApJ, 623, 31.

Eigenbrod, A., et al., 2006, A\&A, 451, 747.

Eliasd\'ottir, A., Hjorth, J., Toft, S., Burud, I., Paraficz, D., 2006, ApJS, 166, 443.

Fadely, R., Keeton, C.R., 2011, AJ, 141, 101.

Fadely, R., Keeton, C.R., 2012, MNRAS, 419, 936.

Fassnacht, C.D., et al., 1999, AJ, 117, 658.

Faure, C., et al., 2008, ApJS, 176, 19.

Faure, C., Sluse, D., Cantale, N., Tewes, M., Courbin, F., Durrer, P., Meylan, G., 2011, A\&A, 536, 29

Floyd, D.J.E., Bate, N.F., Webster, R.L., 2009, MNRAS, 398, 233.

Foreman-Mackey, D., et al., 2013, PASP, 125, 925.

Ghosh, K.K., Narasimha, D., 2009, ApJ, 692, 694.

Hewitt, J.N., Turner, E.L., Schneider, D.P., Burke, B.F., Langston, G.I., 1988, Natur, 333, 537.

Hewitt, J.N., Turner, E.L., Lawrence, C.R., Schneider, D.P., Brody, J.P., 1992, AJ, 104, 968.

Hezaveh, Y.D., et al., 2013, ApJ, 767, 132.

Inada, N., et al., 2003b, AJ, 126, 666.

Inada, N., et al., 2003a, Natur, 426, 810.



Inada, N., et al., 2009, AJ, 137, 4118.

Irwin, M.J., Webster, R.L., Hewett, P.C., Corrigan, R.T., Jedrzejewski, R.I., 1989, AJ, 98, 1989.

Jackson, N., Xanthopoulos, E., Browne, I.W.A., 2000, MNRAS, 311, 389

Jackson, N., 2008, MNRAS, 389, 1311.

Jackson, N., 2011, ApJ, 739, L28.

Jackson, N., 2013, BASI, 41, 19.

Kafle, P.R., Sharma, S., Lewis, G.F., Bland-Hawthorn, J., 3014, ApJ, 794, 59

Kayo, I., et al., 2007, AJ, 134, 1515.

Kayser, R., et al., 1990, ApJ, 364, 15.

Keeton, C.R., Gaudi, B.S., Petters, A.O., 2003, ApJ, 598, 138.

Kellermann, K.I., Sramek, R.A., Schmidt, M., Green, R.F., Shaffer, D.B., 1994, AJ,108, 1163

Kimball, A.E., Kellermann, K.I., Condon, J.J., Ivezic, Z., Perley, R.A., 2011, ApJ, 739L, 29.

Klypin, A., Kravtsov, A.V., Valenzuela, O., Prada, F., 1999, ApJ, 522, 82.


Kochanek, C.S., 2004, in Saas-Fee School: Gravitational Lensing: Strong, Weak and Micro, eds. F. Meylan et al., astro-ph/0407232

Kochanek, C.S., Dalal, N., 2004, ApJ, 610, 69.

Kochanek, C.S., Falco, E.E., Impey, C., Lehar, J., McLeod, B., 
Rix, H.W., 1998. http://www.cfa.harvard.edu/castles/

Kochanek, C.S., et al., 2006, ApJ, 640, 47.

Koopmans, L.V.E., de Bruyn, A.G., 2000, MNRAS, 358, 793

Koopmans, L.V.E., et al., 2003, ApJ, 595, 712

Koopmans, L.V.E., 2005, MNRAS, 363, 1136.

Koposov, S.E., Belokurov, V., Torrealba, G., Evans, N.W., 2015, astro-ph/1503.2079

Kratzer, R.M., et al., 2011, ApJ, 728L, 18.

Laor, A., Behar, E., 2008, MNRAS, 390, 847.

Leipski, C., Falcke, H., Bennert, N., H\"uttemeister, S., 2006, A\&A, 455, 161.

Maccio, A.V., Moore, B., Stadel, J., Diemand, J., 2006, MNRAS, 366, 1529


Mao, S., Schneider, P., 1998, MNRAS, 295, 587.

Metcalf, R.B., Zhao, H., 2002, ApJ, 567, L5.

Metcalf, R.B., 2005, ApJ, 629, 673.

Metcalf, R.B., Amara, A., 2012, MNRAS, 419, 3414.

Minezaki, T., 2007, suba, prop, 90.

Mittal, R., Porcas, R., Wucknitz, O., 2007, A\&A, 447, 515.

Moore, B., et al., 1999, ApJ, 524, L19

Morgan, N.D., Kochanek, C.S., Pevunova, O., Schechter, P.L., 2005, AJ, 129, 2531.

Morgan, C.W., Kochanek, C.S., Dai, X., Morgan, N.D., Falco, E.E., 2008, ApJ, 689, 755.

Mosquera, A., Kochanek, C.S., 2011, ApJ, 738, 96.

Moustakas, L. A., Metcalf, R. B. 2003, MNRAS, 339, 607

Moustakas, L.A., et al., 2012, AAS, 21914601

Mu\~noz, J.A., Mediavilla, E., Kochanek, C.S., Falco, E.E., Mosquera, A.M., 2011, ApJ, 742, 67.

Muxlow, T.W.B., et al., 2005, MNRAS, 358, 1159.

Myers, S.T., et al., 2003, MNRAS, 341, 1

Nierenberg, A. M., Treu, T., Wright, S. A., Fassnacht, C. D., Auger, M. W. 2014, MNRAS, 442, 2434

Ofek, E.O., Maoz, D., Rix, H., Kochanek, C.S., Falco, E.E., 2006, ApJ, 641, 70.

Oguri, M., et al., 2008, MNRAS, 391, 1973.

Ostman, L., Goobar, A., M\"ortsell, E., 2008, A\&A, 485, 403.

Phillips, P.M., et al., 2000, MNRAS, 319L, 7.

Poindexter, S., Morgan, N., Kochanek, C.S., 2008, ApJ, 673, 34.

Ratnatunga, K.U., Griffiths, R.E>, Ostrander, E.J., 1999. AJ, 117, 2010

Riechers, D. A., et. al., 2008, ApJ, 686, 851.

Reimers, D., Hagen, H.-J., Baade, R., Lopez, S., Tytler, D., 2002, A\&A, 382L, 26.

Ricci, D., et al., 2011, A\&A, 528A, 42.

Schinnerer, E., et al., 2007, ApJS, 172, 46.

Schinnerer, E., et al., 2010, ApJS, 188, 384.

Schneider, P., Weiss, A., 1992, A\&A, 260, 1.

Serjeant S., 2012, MNRAS, 424, 2429

Sluse, D., Chantry, V., Magain, P., Courbin, F., Meylan, G., 2012, A\&A, 538A, 99.

Sluse, D., Hutsem\'ekers, D., Courbin, F., Meylan, G., Wambsganss, J., 2012, A\&A, 544A, 62.

Sluse, D., Kishimoto, M., Anguita, T., Wucknitz, O., Wambsganss, J., 2013, A\&A, 553A, 53.

Steenbrugge, K.C., Jolley, E.J.D., Kuncic, Z., Blundell, K.M., 2011, MNRAS, 413, 1735.

Sugai, H., et al., 2007, ApJ, 660, 1016.

Ulvestad, J.S., Wong, D.S., Taylor, G.B., Gallimore, J.F., Mundell, C.G., 2005, AJ, 130, 936

Urry, M., Padovani, P., 1995, PASP, 107, 803

Vegetti, S., Koopmans, L.V.E., 2009, MNRAS, 400, 1583.

Vegetti, S., et al., 2012, Natur, 481, 341.

Walsh, D., Carswell, R.F., Weymann, R.J., 1979, Natur, 279, 381.

Wang, Y., Zhao, H., Mao, S., Rich, R.M., 2012, MNRAS, 427, 1429.

Warren, S.J., Dye, S., 2003, ApJ, 590, 673.

White, R.L., Helfand, D.J., Becker, R.H.; Glikman, E., de Vries, W., 2007, ApJ, 654, 99

Winn, J.N., et al., 2002, AJ, 123, 10.

Wisotzki, L., Koehler, T., Kayser, R., Reimers, D., 1993, A\&A, 278, L15.

Wisotzki, L., Schechter, P.L., Bradt, H.V., Heinm\"uller, J., Reimers, D., 2002, A\&A, 395, 17.

Wisotzki, L., et al., 2004, AN, 325, 135.

Wucknitz, O., Volino, F., 2008, Proceedings of Science (IX EVN Symposium), 102.

Xu, D.D., et al., 2009, MNRAS, 398, 1235.

Xu, D.D., et al., 2015, MNRAS, 447, 3189.

Zackrisson, E., Riehm, T., 2010, AdAst2010E, 9Z, .

Zucker, D.B., et al., 2006, ApJ, 643L, 103

\end{document}